# Optimization of the Superconducting Linear Magnetic Bearing of a Maglev Vehicle

Loïc Quéval[1], Guilherme G. Sotelo[2], Yassin Kharmiz[1], Daniel H. N. Dias[2], Felipe Sass[2],
Víctor M. R. Zermeño[3], Raimund Gottkehaskamp[1]

1. University of Applied Sciences, Düsseldorf, Germany.
2. Fluminense Federal University, Niterói (RJ), Brazil
3. Karlsruhe Institute of Technology (KIT), Eggenstein-Leopoldshafen, Germany
Email : loic.queval@gmail.com

## Abstract

Considering the need for cost/performance prediction and optimization of superconducting maglev vehicles, we develop and validate here a 3D finite element model to simulate superconducting linear magnetic bearings. Then we reduce the 3D model to a 2D model in order to decrease the computing time. This allows us to perform in a reasonable time a stochastic optimization considering the superconductor properties and the vehicle operation. We look for the permanent magnet guideway geometry that minimizes the cost and maximizes the lateral force during a displacement sequence, with a constraint on the minimum levitation force. The displacement sequence reproduces a regular maglev vehicle operation with both vertical and lateral movements. For the sake of comparison, our reference is the SupraTrans prototype bearing. The results of the optimization suggest that the bearing cost could be substantially reduced, while keeping the same performances as the initial design. Alternatively, the performances could be significantly improved for the same original cost.

## I. Introduction

Recently, a full scale superconducting maglev vehicle, named MagLev-Cobra, was successfully demonstrated at the Federal University of Rio de Janeiro, Brazil [1, 2]. In order to achieve the superconducting levitation, it uses a permanent magnet (PM) guideway and several cryostats containing YBCO superconducting bulks placed on the bottom of the vehicle. The interaction of the PM guideway static magnetic field with the superconductors produces stable levitation and lateral guidance forces. One issue of this technology is the cost, because of the large amount of rare earth PM used along the entire line. In this context, the optimization of the superconducting linear magnetic bearing (SLMB) focusing on the reduction of PM guideway material used, while maintaining the performances, motivates this work.

Lately several models have been successfully developed by different research groups around the world to simulate superconducting magnetic bearings [3, 4, 5, 6, 7]. However, their use has been so far generally limited to parametric studies involving few hundred evaluations [8, 9]. Indeed the problem is highly nonlinear and intrinsically time-dependent, leading to very large computing time in order to perform the numbers of simulations needed in a stochastic optimization process. This is the reason why, in [10, 11], the shape of the PM guideway was optimized considering the superconductor as a perfect diamagnet. A SLMB model based on the critical state approximation was only used in a second step for verification.

Besides, in previous works, only vertical motion was considered. But during the regular operation of a maglev vehicle, lateral displacements are to be expected, due to curves for example. This tends to modify the magnetization [12], and as a result to reduce the levitation force and alter the lateral force [13, 14]. This should be taken into account during the optimization.

We propose here to perform a stochastic optimization of the PM guideway geometry, taking both the superconductor properties and the displacement sequence into consideration. To this aim, we start by developing and validating a 3D H-formulation finite element model with a power law E-J relationship. Then we discuss a proper 2D model reduction that allows us to reduce the computing time. Finally, we use the 2D model to perform the stochastic optimization considering a displacement sequence, which reproduces the regular maglev vehicle operation with both lateral and vertical movements.

## II. Modeling and Validation

In this study, we consider the SLMB initially designed and optimized for the SupraTrans maglev vehicle demonstrator [15, 16]. Note that this SLMB was designed considering neither the HTS bulk properties nor the vehicle displacement sequence. The coordinate system adopted in this work is shown in Fig. 1.

### A. PM guideway model

The permanent magnet guideway is made of Nd-Fe-B permanent magnets ($B_r$ = 1 T) and soft ferromagnetic pieces (SAE1010)

arranged in flux concentrator. Its geometry is shown in Fig. 1. It is modeled using a 2D magnetostatic finite element model. It includes the iron nonlinear BH curve and the guideway real geometry. In Fig. 2, the magnetic flux density measured at several distances above the PM guideway is compared with the calculated values.

*B. HTS bulk model*

The superconducting bulk is a 3-seeded melt-textured YBCO block. Its geometry is shown in Fig. 1. It is modeled using an H-formulation finite element model implemented in COMSOL Multiphysics 4.3a PDE mode application [17, 18, 19, 20]. It uses a power law $E$-$J$ relationship, and an isotropic Kim like model [21] to describe the dependence of the critical current density on the magnetic flux density. The importance of modeling each grain of the bulk has been underlined in [22]. Therefore, we model the bulk with 3 superconducting domains, each one having a net current enforced to zero by means of integral constraints. In the following, we use both 2D and 3D HTS bulk models.

*C. Superconducting linear magnetic bearing model*

The superconducting linear magnetic bearing (SLMB) model is built by unidirectional coupling between the PM guideway model and the HTS bulk model. The coupling is done by applying the sum of the external field and the self-field on the outer boundaries of the superconducting bulk model [23]. The external field is obtained from the PM guideway model. We rely here on the hypothesis that the field generated by the supercurrent does not significantly influence the coercive field of the PM. The self-field is calculated at each time step by means of the Biot-Savart law.

Such approach requires (a) only one static solution of the PM guideway model, and (b) a reduced $LN_2$ domain around the HTS bulk [23]. Thanks to this modeling strategy, the model is fast enough to perform optimization in a reasonable time.

*D. Measurements procedure*

In this article, we use experimental data obtained with the SLMB test rig presented in [24, 25]. The experimental protocol is the following. First, the HTS bulk is cooled with liquid nitrogen at 77 K at a given position. Second, the HTS bulk is moved vertically (z-direction) and laterally (y-direction) at a speed of 1 mm/s, while the levitation force $F_z$ and lateral force $F_y$ are measured. We performed two test sequences.

▪ ZFC sequence: (a) The bulk is cooled down at a distance of 100 mm above the center of the PM guideway, where the PM guideway magnetic field can be neglected (zero field cooling). (b) The bulk is moved vertically downward until the gap between the bulk and the PM guideway is 5 mm. (c) The bulk is moved vertically upward to its initial position. This sequence is used for calibration and validation.

▪ LD sequence: (a) The bulk is cooled down at a distance of 25 mm above the center of the PM guideway. (b) The bulk is moved vertically downward until the gap between the bulk and the PM guideway is 5 mm. (c) The bulk is moved laterally with amplitude of 10 mm first to the right, then to the left, and finally back to the center. Such sequence reproduces approximately the regular operation of a maglev vehicle [13]. It is used for validation and optimization.

*E. Validation*

In this section, we use the 3D HTS bulk model. It is common practice to calibrate the parameters of the bulk using the ZFC sequence maximum levitation force [4, 24, 26]. Accordingly, the critical current density $J_{c0}$ is set here at $1.0 \cdot 10^8$ A/m², so that the simulated maximum levitation force during the ZFC sequence is equal to the measured value (Fig. 3). The others HTS bulk parameters are summarized in Table I.

TABLE I
YBCO BULK PARAMETERS

| Symbol | Quantity | Value |
|--------|----------|-------|
| $E_c$ | Critical current criterion | $1 \cdot 10^{-4}$ V/m |
| n | Power law exponent | 21 |
| $B_0$ | Kim model parameter | 0.37 T |
| $\rho_{air}$ | Air resistivity | 100 $\Omega$.m |

The forces measured are in fair agreement with the force calculated with the SLMB model (Figs. 3 and 4). The difference is attributed to the model calibration, the bulk deterioration [24], the spatial inhomogeneity of top-seed melt-textured bulks [27], and the intergrain current [28]. This serves as a validation. Fig. 5 shows the anticipated reduction of the levitation force during LD sequence. Note that the various features of our model permit to obtain a better accuracy, than previously reported [24, 25, 26].

## III. OPTIMIZATION

*A. Model reduction*

Since the computing time achieved with the 3D bulk model is far too long for optimization, we consider two reduced 2D models. The first 2D model is obtained by reducing artificially $J_{c0}$, so that the simulated maximum levitation force during the ZFC sequence is equal to the measured value. The length in the x-direction is taken equal to the physical bulk dimension. This is similar to the 3D model calibration performed above, and to what was done in [24, 25, 26]. The second 2D model is obtained by adopting the same value for $J_{c0}$ as the 3D bulk model. But the length in the x-direction is artificially shortened, so that the simulated and measured maximum levitation force during the ZFC sequence are equal. This is conceptually similar to accounting for an effective length As shown in Figs. 3, 4 and 5, the 2D model using a shorter length approximates better the 3D model, and is therefore used in the following.

*B. Parametrization*

In view of the optimization, we simplify and parametrize the guideway geometry as shown in Fig. 6. To further reduce the computing time of the PM guideway model, we use linear elements, a maximum element size of 2 mm for the guideway domains, and set the outer boundary size to 6 times the PM guideway main dimension. For the superconducting bulk model, we use linear edge elements, a mapped mesh of 6×4 elements in each superconducting domain, and set the $LN_2$ domain thickness to half the minimum levitation gap. This allows us to decrease the computing time of the SLMB model to about 1 min (Intel i7-3770, 3.40 GHz, 4 cores, RAM 16 GB), without significant loss of accuracy.

*C. Objective and constraints*

We look for PM guideways that minimize the price of the PM guideway and maximize the lateral force during LD sequence, with a constraint on the minimum levitation force. The dimensions of the HTS bulk are not modified. The bi-objective optimization problem can be formulated as,

$$minimize\ \big(f_1(a,b,c,d), f_2(a,b,c,d)\big) \qquad (1)$$
$$subject\ to\ (c_1(a,b,c,d) < 0)$$

with,

$$f_1(a,b,c,d) = 2ab \cdot \gamma_{PM} + (2c+d)b \cdot \gamma_{Fe} \qquad (2)$$
$$f_2(a,b,c,d) = -F_y\ (t=50\ s) \qquad (3)$$
$$c_1(a,b,c,d) = \delta - \min(F_z\ (t > 20\ s)) \qquad (4)$$

where the price of the PM material and the iron are arbitrarily set to $\gamma_{PM} = 250$ k€/$m^3$ and $\gamma_{Fe} = 2.5$ k€/$m^3$, respectively. The minimum levitation force $\delta$ is taken equal to 77 N, which corresponds to the minimum force produced by the experimental SLMB for the same sequence (Fig. 5).

*D. Bi-objective optimization algorithm*

For the optimization, we couple the SLMB model with a Multi-Objective Particle Swarm Optimization (MOPSO) algorithm [29]. The weighting factor is set at 0.8. The social and cognitive learning factors are both set at 1. The size of the swarm is set at 100 particles and 25. Leaders are selected from an elitist archive. Total computation time was roughly 40 h.

*E. Results and discussion*

The Pareto optimal solutions are shown in Fig. 7. Note that the initial PM guideway I is a Pareto dominated solution. Three optimal PM guideways are plot in Fig. 7, and summarized in Table II. PM guideway A has the same lateral force and minimum levitation force than the initial PM guideway for the lowest cost (-23 %). PM guideway B has the highest lateral force (+38 %) and the same minimum levitation force than the initial PM for the same cost. PM guideway C shows that one could obtain an even higher lateral force at a higher cost. We underline that the optimization results strongly depend on the relative price of the PM material and of the iron, as well as on the displacement sequence considered. However, giving its speed, our model can easily be used to recalculate new estimates if these parameters and sequences are updated.

TABLE II
OPTIMIZATION RESULTS

| PM guideway | $a$ [mm] | $b$ [mm] | $c$ [mm] | $d$ [mm] | $f_1$ [€/m] | $f_2$ [N] |
|---|---|---|---|---|---|---|
| I | 40 | 50 | 14 | 12 | 1005 | -83 |
| A | 27 | 57 | 14 | 18 | 776 | -83 |
| B | 33 | 61 | 10 | 16 | 1012 | -99 |
| C | 36 | 90 | 9 | 16 | 1628 | -129 |

IV. CONCLUSION

In this work, we developed a superconducting linear magnetic bearing 3D finite element model for a maglev vehicle. It is based on the H-formulation with a power law E-J relationship. The dependence of the superconductor critical current on the magnetic flux density, the PM guideway real geometry and the iron nonlinearity are included. The model has been validated by comparison with experimental data. For the optimization, the 3D model was reduced to a 2D model by shortening artificially its length, instead of decreasing the critical current density. It allowed us to obtain a good level of accuracy while keeping a reasonable computational time. Then, the model was coupled with a multi-objective particle swarm optimization algorithm. Taking the SupraTrans prototype SLMB as reference, the PM guideway optimization results show that it is possible to greatly reduce the cost for the same performances on a given displacement sequence; or to greatly improve the performances for the same cost. This is of particular interest, mainly when considering the construction of lines of several kilometers. Overall, this work demonstrates the importance of numerical modeling in order to bring large scale applications of superconductors from the laboratory scale to the industrial level.

<div align="center">REFERENCES</div>

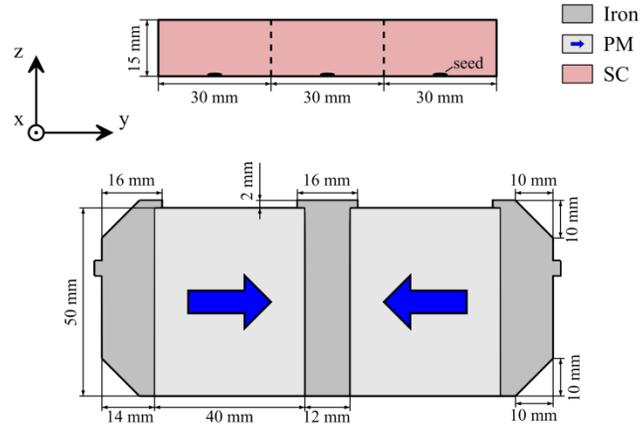

Fig. 1. PM guideway and HTS bulk geometry [24]. The dimension of the HTS bulk in the x-direction is 36 mm. The arrows indicate the PM magnetization direction.

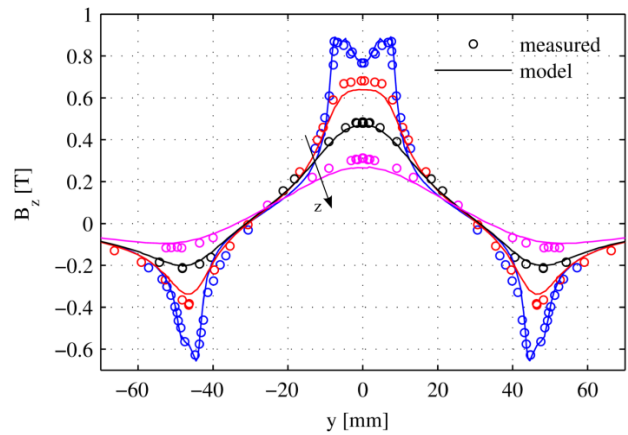

Fig. 2. Magnetic flux density above the PM guideway at z= 1, 5, 10, 20 mm. Measured data from [25].

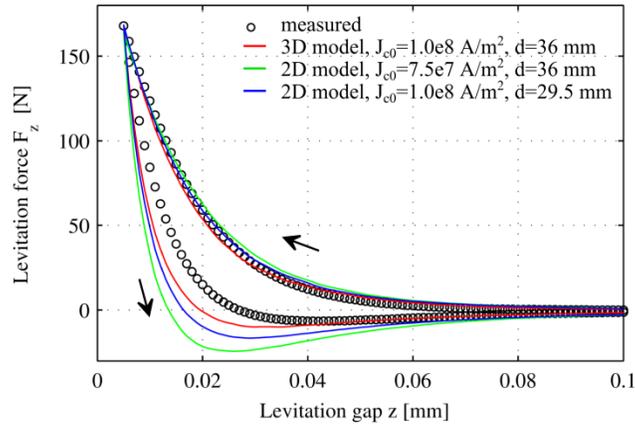

Fig. 3. Levitation force for ZFC sequence. Measured data from [24].

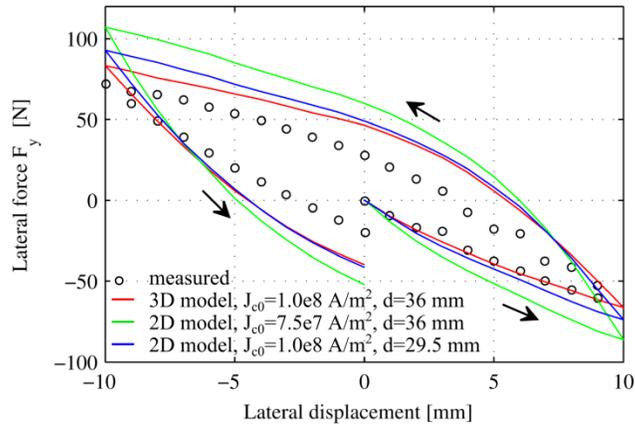

Fig. 4. Lateral force for LD sequence. Measured data from [25].

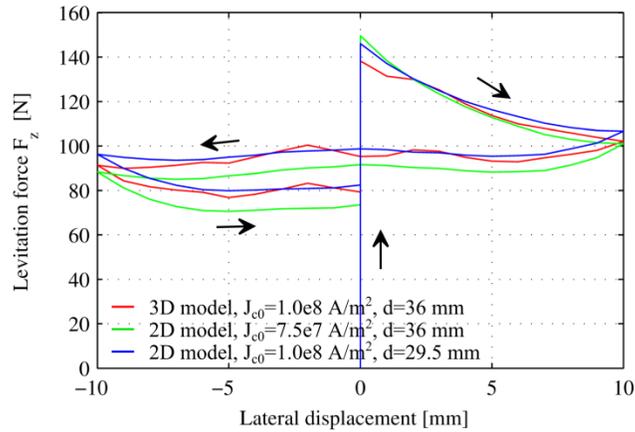

Fig. 5. Levitation force for LD sequence. Measured data unavailable.

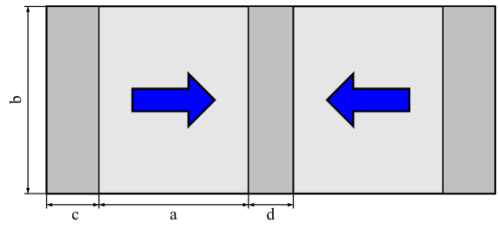

Fig. 6. Parametrization of the PM guideway.

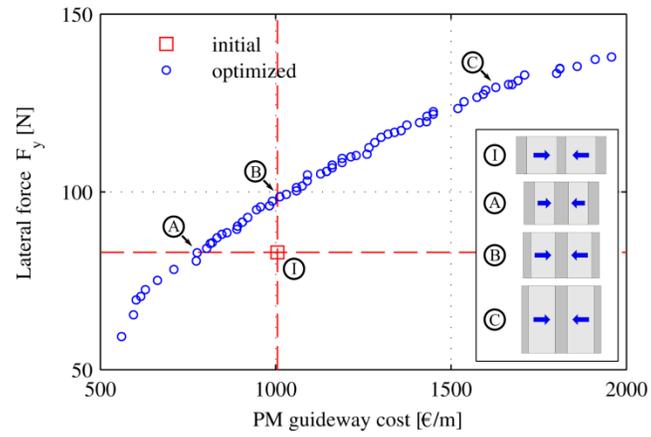

Fig. 7. Pareto optimal solutions of the PM guideway bi-objective optimization. The various PM guideways are drawn on scale.